\documentclass[manuscript,nonacm]{acmart}

\newcommand{\sysname}{\mbox{DimSteer}}
\usepackage{tikz}
\usepackage{amsmath}
\usepackage{float}
\usepackage{booktabs}
\usepackage{amsfonts}
\usetikzlibrary{shapes.geometric, arrows.meta, positioning, fit, backgrounds, calc}

\setcopyright{none}
\renewcommand\footnotetextcopyrightpermission[1]{}
\acmConference{}{}{}
\acmBooktitle{}
\acmDOI{}
\acmISBN{}
\AtBeginDocument{%
  }

\setcopyright{acmlicensed}

\begin{document}

%%
%% The "title" command has an optional parameter,
%% allowing the author to define a "short title" to be used in page headers.
\title{\sysname: Steering LLM Authoring with Automatically Discovered Stylistic Controls}

%%
%% The "author" command and its associated commands are used to define
%% the authors and their affiliations.
%% Of note is the shared affiliation of the first two authors, and the
%% "authornote" and "authornotemark" commands
%% used to denote shared contribution to the research.

%%
%% By default, the full list of authors will be used in the page
%% headers. Often, this list is too long, and will overlap
%% other information printed in the page headers. This command allows
%% the author to define a more concise list
%% of authors' names for this purpose.
\author{Ajit Mallavarapu}
\authornote{Both authors contributed equally to this work.}
\email{am3574@cornell.edu}
\affiliation{%
  \institution{Cornell Tech}
  \country{USA}
}

\author{Ziwei Gu}
\authornotemark[1]
\email{ziweigu@g.harvard.edu}
\affiliation{%
  \institution{Harvard University}
  \country{USA}
}
%%
%% The abstract is a short summary of the work to be presented in the
%% article.

\begin{abstract}
   Large language model writing interfaces often make users steer outputs by repeatedly articulating desired changes in natural language. Yet writers may recognize useful stylistic directions only after seeing alternatives, making revision recall-heavy. We present \sysname{}, an authoring interface that samples prompt-local completions, discovers high-variance activation-space axes of variation, labels them, and exposes them as sliders with pole previews, diff comparison, and reset controls. Users can manipulate discovered dimensions, reducing the need to reformulate prompts for each stylistic adjustment. In a within-subjects study with 16 participants against a matched prompt-only baseline, \sysname{} reduced mental demand, effort, and frustration while preserving comparable perceived success. Participants valued the surfaced dimensions, yet 15 of 16 disagreed that they would have thought to request the same changes in a prompt. Results suggest prompt-local controls can shift LLM authoring from recall-based prompting toward recognition-based exploration and direct manipulation, while preserving prompting for open-ended edits.
\end{abstract}


%% A "teaser" image appears between the author and affiliation
%% information and the body of the document, and typically spans the
%% page.
% \begin{teaserfigure}
%  \includegraphics[width=\textwidth]{sampleteaser}
%   \caption{Seattle Mariners at Spring Training, 2010.}
%   \Description{Enjoying the baseball game from the third-base
%   seats. Ichiro Suzuki preparing to bat.}
%   \label{fig:teaser}
% \end{teaserfigure}

% \received{20 February 2007}
% \received[revised]{12 March 2009}
% \received[accepted]{5 June 2009}

%%
%% This command processes the author and affiliation and title
%% information and builds the first part of the formatted document.
\begin{teaserfigure}
    \centering
    \includegraphics[width=\textwidth]{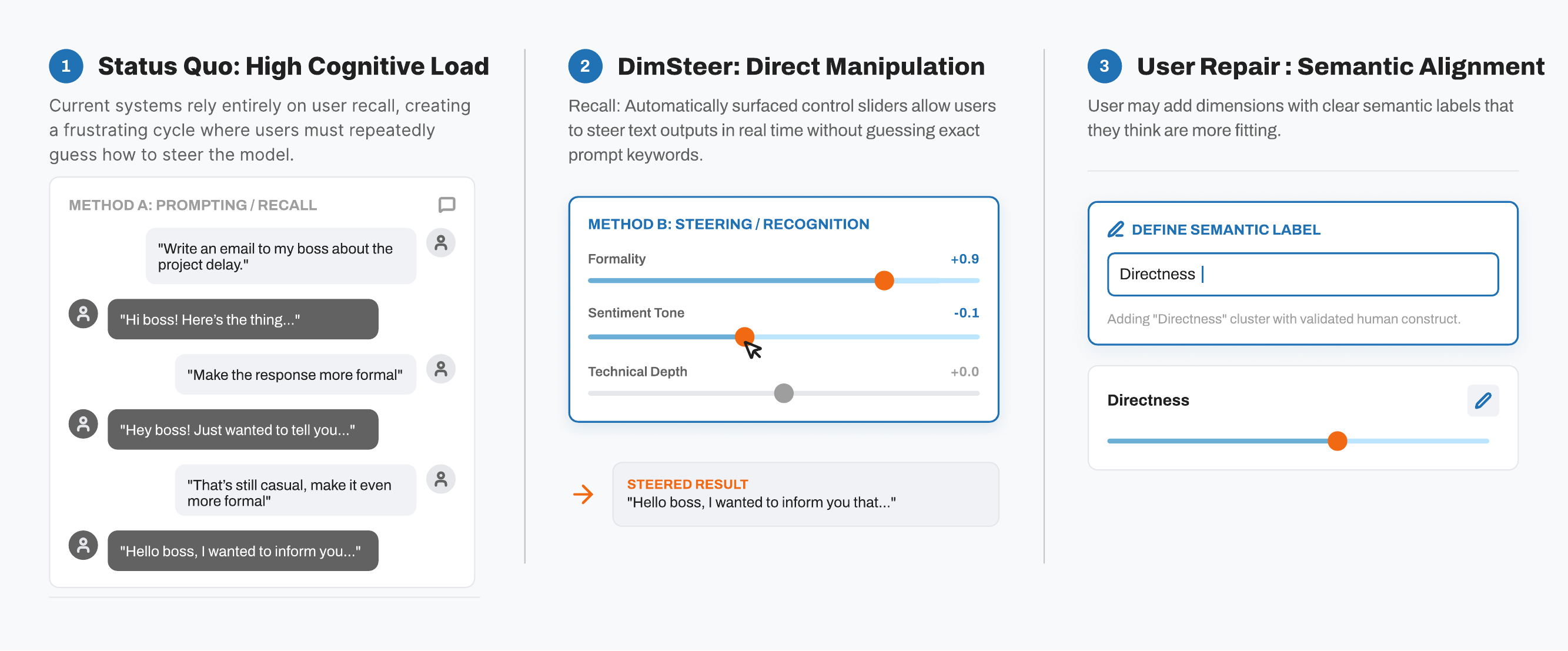}
    \caption{\textbf{\sysname{} shifts LLM interaction from recall-based prompting to direct manipulation.}
    (1) Standard chat interfaces require users to articulate the exact changes needed to steer text.
    (2) \sysname{} instead surfaces automatically discovered dimensions of generation variation as continuous controls, allowing users to explore variation through recognition and direct manipulation.
    (3) When discovered controls are insufficient, users can define custom dimensions to extend the machine-proposed control space.}
    \Description{A three-stage overview of DimSteer showing a transition from recall-based prompting, to automatically discovered semantic sliders, to user-defined repair and extension of the control space.}
    \label{fig:teaser}
\end{teaserfigure}

\maketitle

\section{Introduction}

Writing with a large language model rarely ends at the first draft. The content may be right, yet something about the text is off: the tone too eager, the hedging too heavy, the rhythm not one's own. In chat interfaces, the main way to correct these mismatches is to describe them in language. Prior work shows why this can be difficult: non-experts struggle to formulate effective prompts~\cite{zamfirescu2023johnny}; in-the-wild writing sessions frequently involve revising intents and modifying prior generations~\cite{mysore2025prototypical}; and as alternatives accumulate, comparing outputs and attributing changes become additional sensemaking burdens~\cite{gero2024sensemaking, masson2024directgpt, zhang2026malleable}. The deeper issue is not simply whether a model can produce a desired style, but whether a writer can identify and articulate the right preference precisely enough to steer it~\cite{subramonyam2024envisioning}.

A growing body of work responds by combining language with direct manipulation~\cite{shneiderman1983direct, hutchins1985direct}. DirectGPT brings localized, direct editing to LLM outputs~\cite{masson2024directgpt}; PromptCanvas turns task-relevant facets into dynamic widgets~\cite{amin2025promptcanvas}; and Malleable Prompting reifies preference expressions as sliders, toggles, dropdowns, and other controls~\cite{zhang2026malleable}. Related systems expose semantically defined steering axes in chat~\cite{bo2025steerable} and image generation~\cite{jain2025adaptivesliders}. These approaches make a specified preference easier to manipulate. But the control space is still defined semantically---by what a user says, what a designer chooses, or what a language model proposes---rather than discovered from the model's own prompt-local variation. This distinction matters because preferences are often constructed in context rather than retrieved fully formed~\cite{slovic1995construction}, and interactive elicitation can surface considerations users did not anticipate on their own~\cite{li2023eliciting}. In classic HCI terms, this suggests an opportunity to shift from recall toward recognition~\cite{nielsen1994heuristics}: instead of requiring writers to name every useful dimension before they can manipulate it, an interface could surface candidate dimensions for them to recognize and evaluate.

Rather than asking writers to enumerate the control space, we derive candidate controls from variation already present in the model's responses. Prior interpretability work shows that some high-level model properties admit approximately linear representations and can be manipulated through activation-space interventions~\cite{park2024linear, turner2023activation, zou2025representationengineeringtopdownapproach, rimsky2024steering}, while other features are intrinsically multidimensional~\cite{engels2025linear}. Most recently, researchers sample the model repeatedly and use the resulting activation variance to recover prompt-specific stylistic axes without manually specifying the target attributes, demonstrating that these axes often correspond to stylistic dimensions people later request \cite{mallavarapu2026sampling}. This establishes a technical route to discovering controls. It leaves the central HCI question open: can writers recognize, interpret, and productively use machine-discovered dimensions they did not articulate themselves?

We present \sysname{}, an LLM authoring interface that closes this loop from discovery to control. For each prompt, \sysname{} samples a cloud of completions, extracts the one to three highest-variance stylistic axes from the model's activations, labels them, and renders them as continuous sliders beside the draft, each tagged with the share of variation it explains. Writers steer by dragging: adjustments are applied through activation-level interventions rather than re-prompting, and a persistent diff view shows what changed between versions. When discovery misses a dimension a writer wants, they can define their own control, which is compiled on the fly into a contrastive steering vector~\cite{rimsky2024steering}. The pipeline is training-free and prompt-conditional: the controls presented to a writer are specific to what the model itself varies for that instruction.

We evaluate \sysname{} in a within-subjects study (N = 16) against a matched prompt-only baseline using the same target model and diff view. Prior work has already shown that direct-manipulation controls can improve preference adjustment when the relevant dimensions are specified in advance~\cite{zhang2026malleable}; our central question is what happens when the dimensions themselves are discovered. Participants reported lower mental demand, effort, and frustration with \sysname{} while maintaining comparable perceived success. They also strongly endorsed the surfaced dimensions as useful, even though most reported they would not have thought to request the same changes in a prompt. We examine qualitative cases in which participants recognized useful dimensions after they were surfaced, ranging from relabeling vague prior intentions to discovering conceptual handles they explicitly said they could not name beforehand.
\newline
\newline
This paper makes three main contributions:
\begin{enumerate}
    \item \textbf{\sysname{},} a training-free LLM authoring interface that exposes automatically discovered, prompt-conditional stylistic dimensions as continuous steering controls, closing the loop between representation-level discovery and interactive control.
    \item \textbf{Empirical results from a within-subjects study (N = 16),} showing that prompt-conditional continuous steering reduced perceived mental demand, effort, and frustration compared to natural-language prompting, while surfacing stylistic dimensions that writers wanted but often would not have articulated on their own.
    \item \textbf{An account of how writers interpret, trust, and repair machine-named controls,} with design implications for interfaces that combine representation-level discovery with direct manipulation in generative writing tasks.
\end{enumerate}

\section{Related Work}

\subsection{Prompting and LLM Writing Workflows}

LLM writing interfaces commonly rely on natural-language prompting as the primary mechanism for control. In these interfaces, the prompt acts not only as an instruction to the model, but also as the user's main mechanism for specifying preferences, constraints, and revisions. Prior work has shown that this interaction model can impose substantial cognitive demands. Zamfirescu-Pereira et al. found that non-AI experts often struggle to formulate effective prompts, relying on brittle strategies and folk theories about model behavior~\cite{zamfirescu2023johnny}. Subramonyam et al. describe a related ``gulf of envisioning,'' in which users have difficulty anticipating what a model can produce and translating underspecified goals into effective prompt instructions~\cite{subramonyam2024envisioning}. In writing contexts, this difficulty is amplified because desired changes are often stylistic, rhetorical, or affective rather than clearly specified factual requirements.

Studies of LLM-assisted writing further show that writing with models is iterative rather than one-shot. Mysore et al. identify recurring patterns of human-AI collaboration in which users revise intents, compare alternatives, and modify prior generations over time~\cite{mysore2025prototypical}. Gero et al. similarly emphasize that working with multiple LLM outputs creates sensemaking burdens: users must compare versions, identify differences, and infer how a prompt or model behavior led to a particular output~\cite{gero2024sensemaking}. These findings motivate interfaces that reduce the burden of articulating and tracking stylistic changes. \sysname{} builds on this motivation by shifting part of the interaction from prompt formulation to recognition and manipulation of surfaced dimensions of variation.

\subsection{Direct Manipulation for LLM Interfaces}

The design of \sysname{} is grounded in the HCI principle of direct manipulation: interfaces can reduce cognitive effort by making objects and actions visible, incremental, and reversible~\cite{shneiderman1983direct,hutchins1985direct}. Recent LLM interfaces apply this principle to generative text and media. DirectGPT enables users to manipulate generated text through localized editing operations rather than repeatedly formulating full natural-language prompts~\cite{masson2024directgpt}. PromptCanvas provides a composable workspace in which prompt components and task facets can be represented as dynamic widgets~\cite{amin2025promptcanvas}. Malleable Prompting turns user-authored preference expressions into sliders, toggles, dropdowns, and other controls, allowing users to adjust preferences after they have been expressed in language~\cite{zhang2026malleable}. Dynamic Prompt Middleware similarly generates contextual prompt-refinement controls to help users refine AI responses~\cite{drosos2025dynamic}.

Malleable Prompting is the closest prior work to ours in its goal of turning prompt-level preferences into manipulable interface controls: it takes preferences that users have already expressed in natural language and reifies them as widgets that can be adjusted after the fact~\cite{zhang2026malleable}. \sysname{} differs in the source, timing, and mechanism of control. Its sliders are not derived from user-stated preferences, but from variation the model already exhibits across sampled completions for the current prompt; they are surfaced before the user has named the relevant dimension; and moving a slider steers internal activations rather than rewriting or parameterizing the prompt.

More broadly, prior widget-based systems show that direct manipulation can make specified preferences easier to adjust. In systems such as PromptCanvas and Dynamic Prompt Middleware, controls are generally derived from task context, prompt components, designer choices, or LLM-generated prompt refinements~\cite{amin2025promptcanvas,drosos2025dynamic}. \sysname{} instead studies whether writers can recognize and use automatically discovered, prompt-local controls during revision.

Semantic sliders have also been explored in other generative domains. AdaptiveSliders, for example, supports user-aligned slider-based editing for text-to-image generation~\cite{jain2025adaptivesliders}. Like these systems, \sysname{} uses continuous controls to support exploration. Unlike them, it focuses on LLM authoring and studies how writers interpret and use automatically discovered prompt-local controls during text revision.

\subsection{Preference Elicitation and Recognition over Recall}

\sysname{} also builds on work showing that user preferences are often constructed through interaction rather than retrieved fully formed. Slovic argues that preferences are frequently assembled in context, shaped by the options and tradeoffs people encounter~\cite{slovic1995construction}. Related work on preference elicitation shows that interactive systems can surface considerations users did not initially articulate~\cite{li2023eliciting}. This perspective is especially relevant for LLM authoring: a writer may not know in advance that they want a message to be more direct, warmer, less deferential, or more vivid, but may recognize that preference once the dimension is made visible.

This connects to the classic HCI principle of recognition over recall~\cite{nielsen1994heuristics}. Prompt-only interfaces require users to recall and formulate the relevant revision instruction before they can act on it. \sysname{} instead presents candidate dimensions of variation as inspectable controls. The central question is therefore not only whether users can manipulate known preferences more easily, but whether automatically surfaced dimensions help them recognize preferences they had not explicitly named.

\subsection{Activation Steering and Representation-Level Control}

\sysname{} uses activation-space interventions as the mechanism for applying discovered controls. Prior work has shown that model behavior can be influenced by modifying internal representations without updating model weights. Activation Addition demonstrates that adding activation differences associated with natural-language prompts can steer model behavior~\cite{turner2023activation}. Representation Engineering frames internal representations as manipulable objects for transparency and control~\cite{zou2025representationengineeringtopdownapproach}. Contrastive Activation Addition constructs steering directions from paired examples that contrast a desired behavior with an undesired one~\cite{rimsky2024steering}. Recent steerable chat systems further show that activation steering can support interactive user-facing control~\cite{bo2025steerable}.
 
Most activation-steering methods begin with a target concept specified by a researcher, developer, or user, such as sentiment, truthfulness, refusal, or another 
named behavior~\cite{turner2023activation,zou2025representationengineeringtopdownapproach,rimsky2024steering}. \sysname{} differs by using the model's own prompt-local variation to propose candidate controls. For a given prompt, the system samples multiple completions, extracts internal activations, performs PCA over the resulting variation cloud, and surfaces the top axes as interface controls. This makes the system less dependent on a predefined vocabulary of controllable attributes.

At the same time, discovered axes should not be treated as ground-truth semantic variables. Prior work on the linear representation hypothesis suggests that some model properties can be represented and manipulated along approximately linear directions~\cite{park2024linear}, while other work shows that language-model features are not always one-dimensionally linear or disentangled~\cite{engels2025linear}. Farquhar et al. caution that unsupervised representation-discovery methods can recover salient features without uniquely identifying the intended semantic property~\cite{farquhar2023challenges}. Sparse autoencoder methods offer one route toward more localized and interpretable features~\cite{cunningham2023sparse}, but typically require training a feature dictionary and selecting which features matter for a task.

\sysname{} therefore treats each discovered direction as an \emph{interface hypothesis}: a candidate control that is prompt-local, approximate, and subject to user interpretation. The interface pairs labels with pole previews, diff views, variance tags, reset controls, and custom dimensions so that users can inspect, test, ignore, or repair discovered controls during authoring.

\section{\sysname{}}

\sysname{} is an LLM authoring interface that surfaces automatically discovered dimensions of generation variation as interactive slider controls. Designed for revision-oriented writing tasks, \sysname{} enables users to fine-tune stylistic, rhetorical, or tonal properties of an existing draft through direct manipulation rather than repeated prompt engineering.

The central design premise of \sysname{} is that discovered controls represent \emph{interface hypotheses}—prompt-local handles on observed patterns of variation among candidate generations. Because these handles may occasionally be mislabeled, redundant, or subtle, the interface provides continuous visual feedback, inspection tools, and repair pathways.

\begin{figure}[H]
    \centering
    \includegraphics[width=1\linewidth]{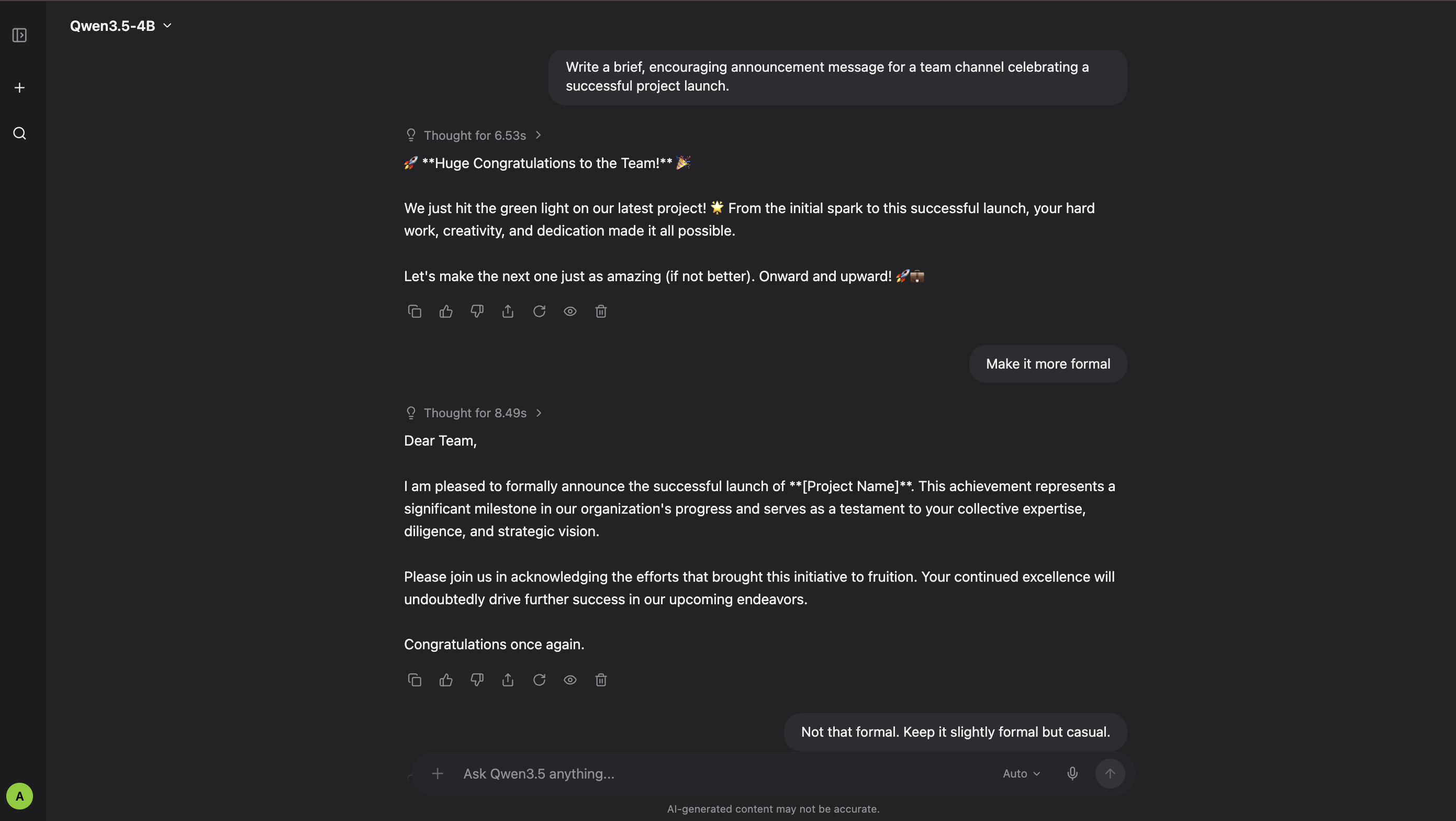} 
    \caption{\textbf{Interface A: Prompt-Only Baseline.} Participants revised drafts using standard natural-language instructions.}
    \label{fig:interface_A}
\end{figure}

\begin{figure}[H]
    \centering
    \includegraphics[width=1\linewidth]{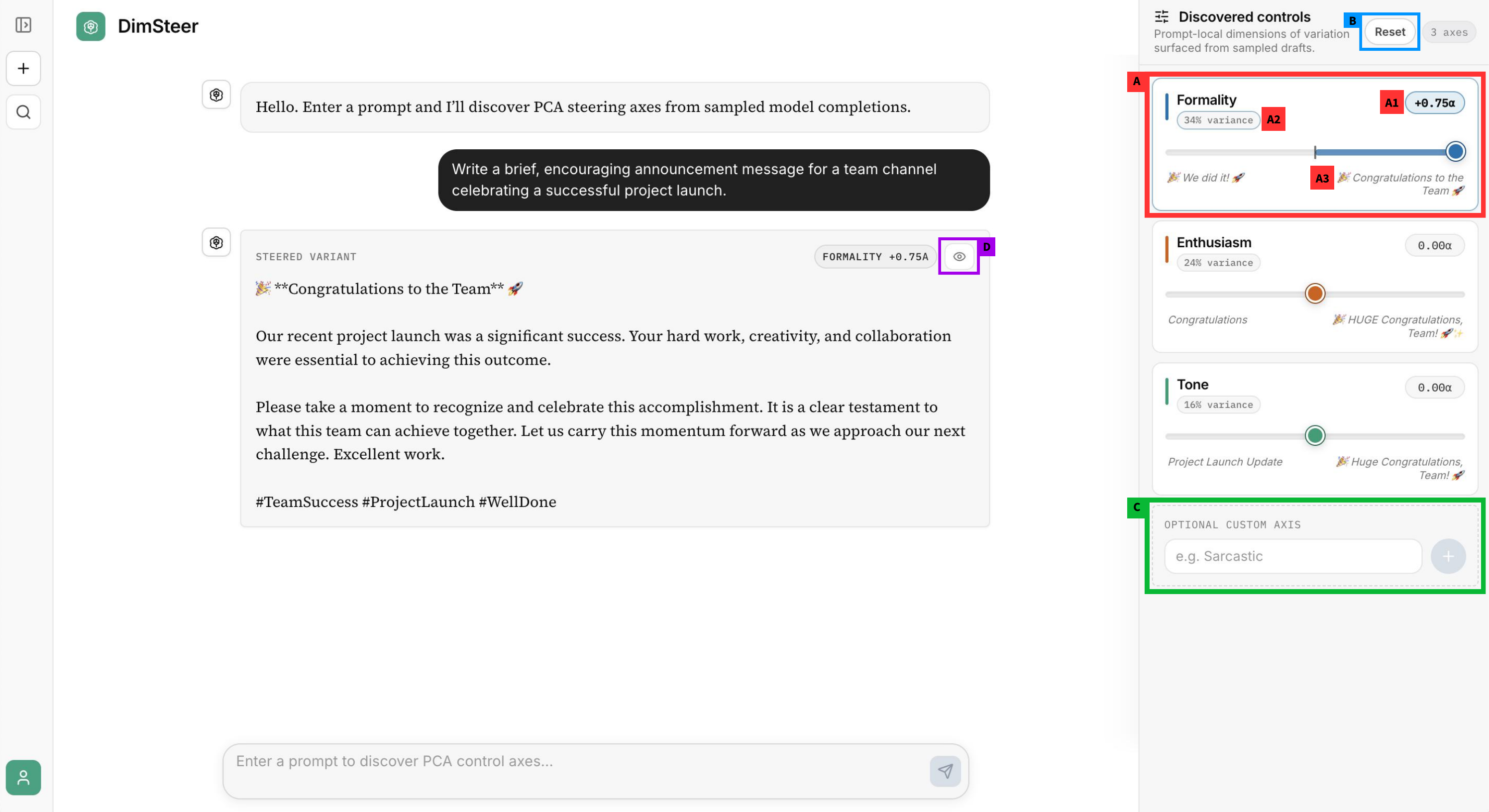}
    \caption{\textbf{The \sysname{} Interface.} (A) Each discovered slider card includes (A1) current steering coefficient, (A2) PCA variance tag, and (A3) directional pole text previews. (B) Reset returns steering to the baseline. (C) Custom-dimension field lets users add missing controls. (D) Diff toggle reveals changes from the baseline draft.}
    \label{fig:interface_B}
\end{figure}

\subsection{Design Goals}

The design of \sysname{} responds to a recurring tension in LLM authoring: natural language is expressive, but users may not know how to name the stylistic change they want until they see alternatives. We therefore drew on HCI work on recognition over recall, direct manipulation, sensemaking, and recoverable human-AI interaction to design controls that are visible and easy to test, but still provisional and correctable.

\paragraph{G1: Support recognition over recall.}
Prompt-based revision asks writers to recall and formulate the right instruction before they can act. However, HCI heuristics emphasize that interfaces can reduce cognitive burden by making relevant actions visible rather than requiring users to retrieve them from memory~\cite{nielsen1994heuristics}. This is especially important when preferences are constructed through interaction rather than fully known in advance~\cite{slovic1995construction,li2023eliciting}. \sysname{} therefore surfaces candidate dimensions of variation directly, so writers can recognize potentially useful directions without first articulating them as prompt instructions.

\paragraph{G2: Enable continuous stylistic adjustment.}
Many writing preferences are matters of degree: a draft can be slightly more formal, much more concise, or somewhat warmer. Direct manipulation interfaces support this kind of incremental control by making operations visible, continuous, and reversible~\cite{shneiderman1983direct,hutchins1985direct}. \sysname{} represents discovered dimensions as sliders so users can explore graded stylistic changes without committing to a full prompt rewrite at each step.

\paragraph{G3: Preserve comparison across iterations.}
LLM authoring often involves evaluating several plausible versions rather than choosing between clearly right and wrong outputs. Prior work shows that comparing generated alternatives creates sensemaking burdens, as users must track what changed and infer which action caused the change~\cite{gero2024sensemaking}. \sysname{} therefore includes diff-based comparison so users can inspect the textual effect of each adjustment and calibrate whether a discovered control behaved as expected.

\paragraph{G4: Support recovery from imperfect controls.}
Because \sysname{}'s controls are automatically discovered, they may be mislabeled, entangled, redundant, or irrelevant to a user's goal. Human-AI interaction guidelines emphasize that users should be able to understand, correct, and recover from imperfect system behavior~\cite{amershi2019guidelines}. \sysname{} therefore treats discovered controls as provisional: users can ignore them, reset them, reinterpret them through pole previews and diffs, or supplement them with manually specified custom dimensions.  

\subsection{Authoring Workflow}

A typical session begins with an initial draft and a panel of discovered controls, each initialized to a neutral baseline. Users inspect the draft, review slider labels and pole previews, adjust one or more controls, and compare the revised draft against the baseline using the diff view. They can then continue adjusting, reset controls, use prompt-based instructions, define a custom dimension, or accept the draft.

This workflow complements natural-language prompting: instruction-based rewrites handle structural or factual updates, while sliders facilitate rapid exploration of stylistic and rhetorical nuance.

\subsection{Discovered Slider Cards \& Pole Previews}
Rather than requiring users to manually name controls, \sysname{} automatically extracts dimensions based on how the model's own outputs vary for the active prompt. 

Each discovered slider card combines a short dimension label, a centered bidirectional slider, a qualitative variance tag, and pole previews showing examples from each end of the axis. The variance tag indicates how prominent the component was in the sampled variation cloud; it is not a confidence score. Pole previews help users interpret ambiguous labels by showing the concrete textual effect of the control for the current prompt.

\subsection{Draft-Level Steering \& Diff-Based Comparison}
When sliders are adjusted, \sysname{} applies global activation interventions to the draft generation. Adjusting multiple sliders simultaneously blends the respective steering directions, enabling writers to navigate stylistic trade-offs (e.g., increasing \emph{Warmth} while decreasing \emph{Length}).

To reduce cognitive load during evaluation, \sysname{} features a diff view highlighting added, deleted, and modified text. This view aids general editing while enabling users to verify whether a slider intervention produced the intended outcome without introducing unintended side effects.

\subsection{Recovery Pathways \& Matched Baseline}
To handle scenarios where automated discovery yields insufficient or noisy dimensions, \sysname{} supports three recovery pathways:
\begin{enumerate}
    \item \textbf{Bypassing/Resetting:} Irrelevant controls can be ignored or reset to zero with one click.
    \item \textbf{Direct Prompting:} Users can issue natural-language edit instructions alongside active slider states.
    \item \textbf{Custom Controls:} Writers can manually enter a target attribute (Figure~\ref{fig:interface_B}C) to synthesize an on-demand steering vector.
\end{enumerate}

\paragraph{Matched Baseline System.} We implemented a prompt-only baseline using the same target model backend, generation parameters, and diff visualization as \sysname{}, but without discovered slider controls (Figure~\ref{fig:interface_A}).

\section{Implementation}
\label{sec:Implementation}

To implement the \sysname{} authoring workflow, we developed a custom full-stack architecture that connects frontend interaction with hidden-state extraction and activation steering. The system has two stages: a dimension-discovery and labeling phase for deriving prompt-local axes of generation variation, and an online steering stage that applies slider-controlled activation interventions during generation.

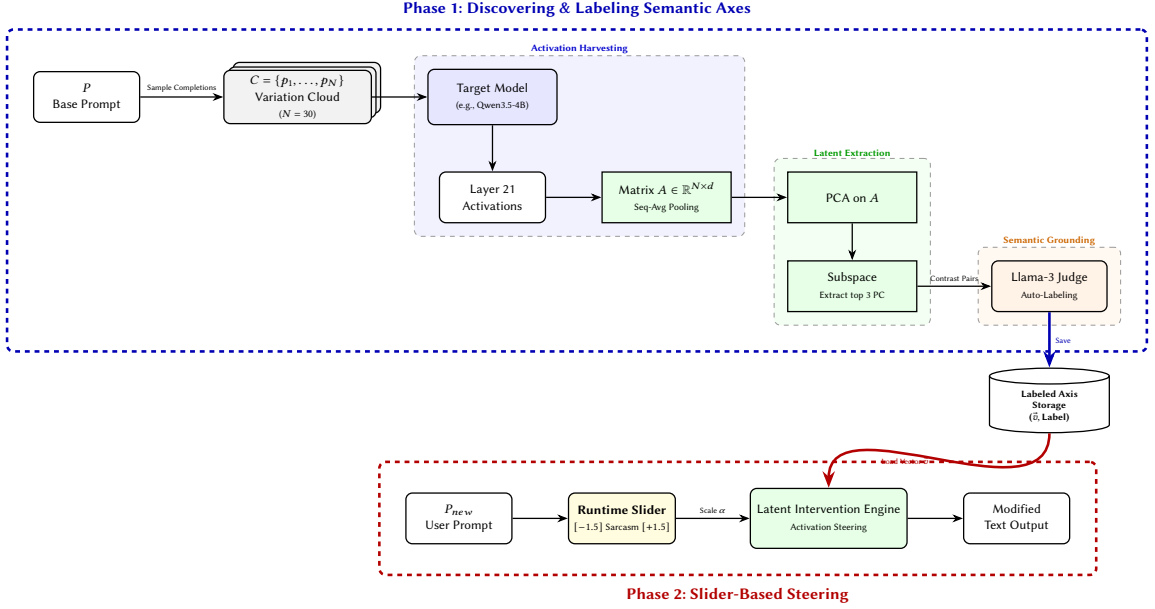
\begin{figure*}[htbp]
    \centering
    \colorlet{OfflineColor}{blue!70!black}
    \colorlet{OnlineColor}{red!70!black}
    
    % Note: Ensure \usetikzlibrary{positioning, fit, backgrounds, arrows.meta, shapes.geometric} is in your document preamble.
    \resizebox{\textwidth}{!}{%
    \begin{tikzpicture}[
        >=Stealth,
        node distance=1.2cm and 1.4cm,
        % Node Styles (widths slightly increased to prevent text overflow)
        basebox/.style={draw, rounded corners, align=center, fill=white, thick, minimum height=1.1cm, minimum width=2.3cm, font=\sffamily\small},
        cloudbox/.style={draw, rounded corners, align=center, fill=gray!10, thick, minimum height=1.1cm, minimum width=3.2cm, font=\sffamily\small},
        modelblock/.style={draw, rectangle, rounded corners, align=center, fill=blue!10, thick, minimum height=1.2cm, minimum width=2.8cm, font=\sffamily\small},
        mathblock/.style={draw, rectangle, align=center, fill=green!10, thick, minimum height=1.1cm, minimum width=2.8cm, font=\sffamily\small},
        judgeblock/.style={draw, rectangle, rounded corners, align=center, fill=orange!10, thick, minimum height=1.1cm, minimum width=2.5cm, font=\sffamily\small},
        storage/.style={draw, cylinder, cylinder body fill=gray!10, cylinder end fill=gray!25, thick, minimum height=1.4cm, minimum width=2.6cm, aspect=0.25, align=center, shape border rotate=90, font=\sffamily\scriptsize\bfseries, inner sep=2pt},
        % Grouping Styles
        grouphighlight/.style={draw=black!40, dashed, rounded corners, inner sep=8pt},
        phasebox/.style={draw, dashed, rounded corners, ultra thick, inner sep=16pt}
    ]

    % ==========================================
    % PHASE 1: Dimension Discovery and Labeling
    % ==========================================
    \node[basebox] (prompt_off) {$P$ \\ Base Prompt};
    
    % 3D Stacked effect for clouds using both xshift and yshift
    \node[cloudbox, right=1.8cm of prompt_off, yshift=0.2cm, xshift=0.2cm] (v2_off) {};
    \node[cloudbox, right=1.8cm of prompt_off, yshift=0.1cm, xshift=0.1cm] (v1_off) {};
    \node[cloudbox, right=1.8cm of prompt_off] (cloud_off) {$C = \{p_1, \dots, p_N\}$ \\ Variation Cloud \\ \scriptsize ($N=30$)};
    
    \draw[->, thick] (prompt_off) -- node[above, font=\sffamily\tiny] {Sample Completions} (cloud_off);

    \node[modelblock, right=1.2cm of cloud_off] (target_off) {Target Model \\ \scriptsize (e.g., Qwen3.5-4B)};
    \node[basebox, below=1.0cm of target_off] (layer_off) {Layer 21 \\ Activations};
    \node[mathblock, right=1.2cm of layer_off] (matrix_off) {Matrix $A \in \mathbb{R}^{N \times d}$ \\ \scriptsize Seq-Avg Pooling};
    
    \draw[->, thick] (cloud_off) -- (target_off);
    \draw[->, thick] (target_off) -- (layer_off);
    \draw[->, thick] (layer_off) -- (matrix_off);

    \node[mathblock, right=1.2cm of matrix_off] (pca_off) {PCA on $A$};
    \node[mathblock, below=0.8cm of pca_off] (gs_off) {Subspace \\ \scriptsize Extract top 3 PC};
    
    \draw[->, thick] (matrix_off) -- (pca_off);
    \draw[->, thick] (pca_off) -- (gs_off);

    \node[judgeblock, right=1.6cm of gs_off] (judge_off) {Llama-3 Judge \\ \scriptsize Auto-Labeling};
    \draw[->, thick] (gs_off) -- node[above, font=\sffamily\tiny] {Contrast Pairs} (judge_off);

    % ==========================================
    % STORAGE BRIDGE
    % ==========================================
    % Positioned perfectly between Phase 1 (above) and Phase 2 (below)
    \node[storage, below=1.2cm of judge_off] (storage) {Labeled Axis \\ Storage \\ \scriptsize ($\vec{v}, \text{Label}$)};
    \draw[->, OfflineColor, ultra thick] (judge_off) -- node[right, font=\sffamily\tiny] {Save} (storage);

    % ==========================================
    % PHASE 2: ONLINE INFERENCE ENGINE
    % ==========================================
    % Anchoring the Engine below and to the left of the storage
    \node[modelblock, fill=green!10, below left=1.2cm and 2.5cm of storage, minimum width=3.4cm, minimum height=1.3cm] (engine) {Latent Intervention Engine \\ \scriptsize Activation Steering};
    
    \node[basebox, fill=yellow!15, left=1.6cm of engine] (slider) {\textbf{Runtime Slider} \\ \scriptsize $[-1.5]$ Sarcasm $[+1.5]$};
    \node[basebox, left=1.2cm of slider] (runtime_prompt) {$P_{new}$ \\ User Prompt};
    \node[basebox, right=1.2cm of engine] (output) {Modified \\ Text Output};

    % Phase 2 Connections
    \draw[->, thick] (runtime_prompt) -- (slider);
    \draw[->, thick] (slider) -- node[above, font=\sffamily\tiny] {Scale $\alpha$} (engine);
    
    % Elegant S-curve from storage down to the engine
    \draw[->, OnlineColor, ultra thick] (storage.south) to[out=270, in=90] node[left=0.1cm, font=\sffamily\tiny] {Load Vector $\vec{v}$} (engine.north);
    
    \draw[->, thick] (engine) -- (output);

    % ==========================================
    % BACKGROUND GROUPS & PHASE BOXES
    % ==========================================
    \begin{scope}[on background layer]
        % Inner Feature Groups
        \node[grouphighlight, fill=blue!5, fit=(target_off)(layer_off)(matrix_off), label={[font=\sffamily\bfseries\scriptsize, text=blue!80!black, yshift=-2pt]above:Activation Harvesting}] (group1) {};
        \node[grouphighlight, fill=green!5, fit=(pca_off)(gs_off), label={[font=\sffamily\bfseries\scriptsize, text=green!60!black, yshift=-2pt]above:Latent Extraction}] (group2) {};
        \node[grouphighlight, fill=orange!5, fit=(judge_off), label={[font=\sffamily\bfseries\scriptsize, text=orange!80!black, yshift=-2pt]above:Semantic Grounding}] (group3) {};
        
        % Outer Phase 1 Box (Encloses all of Phase 1 + the newly named subgroup highlights to guarantee it encapsulates everything)
        \node[phasebox, draw=OfflineColor, fit=(prompt_off)(v2_off)(cloud_off)(group1)(group2)(group3), label={[font=\sffamily\bfseries\large, text=OfflineColor, yshift=4pt]above:Phase 1: Discovering \& Labeling Semantic Axes}] {};

        % Outer Phase 2 Box (Online)
        \node[phasebox, draw=OnlineColor, fit=(runtime_prompt)(slider)(engine)(output), label={[font=\sffamily\bfseries\large, text=OnlineColor, yshift=-4pt]below:Phase 2: Slider-Based Steering}] {};
    \end{scope}

    \end{tikzpicture}%
    }
    \caption{\textbf{\sysname{} implementation pipeline.} The system samples prompt-local completions, extracts layer-21 activations, identifies PCA axes, labels and orients them with an LLM judge, and applies the resulting vectors through slider-based activation steering.}
    \label{fig:restructured_architecture}
\end{figure*}

\subsection{System Architecture \& Model Hosting}

The prototype used a React/Vite frontend connected to a live Python/FastAPI backend running in Google Colab on an NVIDIA T4 GPU. The Colab backend hosted Qwen3.5-4B for generation and steering, exposed REST-style endpoints for prompt generation, axis discovery, and steering, and was made reachable to the local frontend through a LocalTunnel URL. Both the prompt-only baseline and \sysname{} used this same backend and target model, so differences between conditions reflect the interaction mechanism rather than model capability.

The frontend sent user prompts to the backend, displayed returned drafts and discovered controls, and issued steering requests when users adjusted sliders. The backend performed hidden-state extraction, PCA, labeling, polarity alignment, duplicate-axis filtering, and activation steering. Automatic dimension labeling and polarity alignment used Meta-Llama-3-8B-Instruct. Because the study used the live backend, participants experienced actual end-to-end latency, and task-duration logs include these delays. Initial draft generation averaged 13.70s, 18.51s, 8.36s, and 5.61s for shared prompts P1--P4, respectively, while participant-specific open-ended prompts averaged 12.59s. In $C_{STEER}$, axis discovery averaged 8.82s, slider steering averaged 14.90s, and custom-dimension creation averaged 19.43s.

\subsection{Activation Harvesting \& PCA Engine}
For each user prompt, \sysname{} samples $N=30$ stochastic completions from the target model to form a prompt-local variation cloud. For each completion, we extract hidden states from transformer layer 21 over the generated-token window and mean-pool them into a single completion-level activation vector. This produces an activation matrix $A \in \mathbb{R}^{N \times d}$, where each row summarizes one sampled completion.

We mean-center this matrix and apply Principal Component Analysis (PCA) across completions, retaining the top three components as candidate prompt-local steering axes before labeling and deduplication. The variance explained by each component is surfaced in the interface as a descriptive prominence cue, not as a confidence score or guarantee of semantic usefulness. These axes should therefore be interpreted as approximate interface controls rather than ground-truth semantic factors. This discovery procedure is motivated by prior validation of the same sampling-based PCA approach on Qwen3.5-4B~\cite{mallavarapu2026sampling}. In that evaluation, eight annotators provided 245 unconstrained stylistic-control requests across 20 prompts; the top two discovered axes achieved 72.8\% precision and 43.6\% macro-recall against these human-requested dimensions at a semantic matching threshold of $\tau=0.65$. A separate polar-validity evaluation found that 75.6\% of human ratings judged the polar generations accurate to their labels, with 90.9\% adjacent agreement. We use these results as evidence that prompt-local activation variance can surface human-interpretable candidate controls; in this paper, we study how such controls function in a live authoring interface. Sampling parameters and layer-selection details are provided in Appendices~\ref{app:sampling-layer} and~\ref{app:layer-sweep}.

\subsection{Auto-Labeling \& Polarity Alignment Pipeline}
Because the sign of a PCA direction is mathematically arbitrary, the system must ground each axis in representative text to make the directions semantically meaningful to the user.

For each principal component, we project the $N$ sampled generations onto the axis and sort them by projection score. We select exemplar generations from the positive and negative ends of this projection distribution and pass these pole examples to Meta-Llama-3-8B-Instruct in a single structured judging prompt. The judge compares the two poles and returns a structured JSON-style object with a concise label, a polarity judgment, a coherence score, and a short rationale:                       
   \begin{verbatim}
   {
     "label": "<one- or two-word label>",
     "positive_pole_has_more_label": <true or false>,
     "coherence": <numeric score>,
     "rationale": "<one short sentence>"
   }
   \end{verbatim}
The \texttt{label} field is used as the slider name. The \texttt{positive\_pole\_has\_more\_label} field resolves the arbitrary sign of the PCA direction: if it is false, we flip the axis vector and swap the pole examples so that positive slider movement corresponds to increasing the labeled property. The coherence score and rationale are retained as lightweight diagnostics for whether the pole contrast was interpretable.

\subsection{Axis Deduplication}
To reduce redundant controls, \sysname{} filters near-duplicate axis labels after labeling and polarity alignment. Labels are normalized by lowercasing and removing whitespace and punctuation; if two labels share a contiguous character sequence of at least four characters, the system keeps the higher-variance axis and removes the lower-variance duplicate. This lightweight heuristic prevents the interface from showing multiple similarly named sliders that offer little additional user value. 

\subsection{Custom Dimension Creation}
Custom dimensions used Contrastive Activation Addition (CAA) rather than PCA. Given a user-entered attribute, Meta-Llama-3-8B-Instruct helped operationalize the request by generating positive and negative contrastive descriptions for the current prompt. The backend ran Qwen3.5-4B on these contrastive conditions, extracted transformer layer-21 activations over the generated-token window, and used the normalized mean positive-minus-negative activation difference as the steering vector. The resulting CAA direction was exposed through the same slider interface and coefficient range $\alpha \in [-1.5,+1.5]$, providing a repair path when PCA-discovered axes did not cover a desired control.

\subsection{Interactive Steering Mechanism}

When a user moves a slider, \sysname{} applies an activation-space intervention during generation. Each discovered slider corresponds to one prompt-local PCA direction, while each custom slider corresponds to a CAA direction constructed from the user-specified attribute. In both cases, the slider value determines a steering coefficient $\alpha_k$, applied to a unit-norm steering direction $\hat{v}_k$.

During decoding, \sysname{} applies the steering directions at transformer layer 21 to the current decoding position. We selected layer 21 as a fixed mid-to-late intervention layer based on a pilot sweep over candidate layers. The sweep suggested that layers 18--24 formed a usable intervention band, producing visible steering while preserving fluency and task adherence; layer 21 was chosen as a representative midpoint within this band. We used a single fixed layer rather than multi-layer steering to keep the intervention low-parameter and reproducible in the live prototype. Appendix~\ref{app:layer-sweep} provides details. For each generated token $t$, the hidden state $h_{\ell,t}$ is modified as:
   \begin{equation}
   h_{\ell,t}' = h_{\ell,t} + \|h_{\ell,t}\|_2 \sum_{k=1}^{K} \alpha_k \hat{v}_k ,                   
   \end{equation} 
where $\hat{v}_k$ is the unit-norm steering direction for slider $k$ (PCA-discovered or CAA-derived), $\alpha_k$ is the user-controlled coefficient, and $\|h_{\ell,t}\|_2$ scales the intervention to the baseline activation magnitude. Multiple active sliders are combined linearly through the summation over $k$. 

\subsection{Frontend Engineering \& Interface Details}
The frontend is built using React and Vite, focusing on a seamless transition between prompt drafting and slider manipulation. Because each unseen steering configuration requires a backend inference request, rapid slider movements could overwhelm the backend. Slider movements were debounced before being sent to the FastAPI backend, reducing redundant steering requests during rapid interaction. The frontend also cached completed steering results within a session: if a participant returned to a previously viewed slider configuration, the interface restored the cached draft rather than issuing another backend request. Unseen slider configurations still triggered live steering requests.

The UI presents each discovered dimension as a centered continuous slider with steering coefficient $\alpha \in [-1.5, +1.5]$. The center position maps to $\alpha = 0$, corresponding to the unsteered baseline. Moving the slider left applies negative steering toward the negative pole of the discovered axis, while moving it right applies positive steering toward the positive pole. When multiple sliders are adjusted simultaneously, the backend computes a linear combination of the corresponding steering directions.

\section{User Study} 

To evaluate \sysname{}, we conducted a within-subjects laboratory study comparing auto-discovered continuous steering controls against a standard natural-language prompting baseline. The evaluation focused on three questions: whether \sysname{} reduces perceived effort while preserving participants' ability to shape generated text (RQ1), whether auto-discovered controls surface desired dimensions that users may not have articulated themselves (RQ2), and how slider-based steering affects participants' sense of control, exploration, and preference relative to open-ended prompting (RQ3). The user study was conducted under an approved IRB.

\subsection{Participants and Apparatus}  

We recruited 16 participants: 8 female and 8 male. Ages ranged from 18 to 54: 11 were 18--24, 4 were 25--34, and 1 was 45--54. All participants reported prior experience with writing prompts for AI tools: 6 were somewhat experienced, 6 were experienced, and 4 were very experienced. Participants also reported frequent use of AI writing tools: 13 used AI tools for writing or editing daily, and 3 used them a few times per week. All 16 participants completed all study tasks. Participants were compensated \$20 for completing the study.

The study compared two interfaces using the same live backend and the same underlying model, Qwen3.5-4B. Interface A was a standard chat-based prompting interface in which participants revised drafts through natural-language instructions. Interface B was \sysname{}, which presented auto-discovered continuous controls for modifying the generated draft along prompt-local dimensions of variation. Because both interfaces used the same model and backend, differences between conditions reflect the interaction mechanism rather than differences in model capability.  

The apparatus consisted of a local React/Vite frontend connected to a Python/FastAPI backend running in Google Colab. The backend was exposed to the frontend through a LocalTunnel URL. The backend supported prompt-local sampling, activation extraction, PCA, labeling, polarity alignment, duplicate-axis filtering, and activation steering. During the study, the four shared prompts used the same backend-discovered axes across participants to hold the control set constant, while participant-specific prompts used axes generated for that participant's prompt. Slider adjustments then triggered live steering requests to the same backend.

Each participant completed the same set of tasks in both interfaces: four shared prompts covering descriptive writing and workplace communication, plus one participant-specific open-ended prompt. The participant-specific prompt was held constant across both interfaces for that participant. This design allowed us to compare interaction mechanisms within participants while keeping the writing goal fixed across systems.

\subsection{Study Design and Tasks}

The study used a within-subjects design with two counterbalanced conditions:

\begin{itemize}
   \item \textbf{$\mathbf{C_{PROMPT}}$ (Interface A):} Participants refined outputs by typing natural-language instructions into a chat interface.
   
   \item \textbf{$\mathbf{C_{STEER}}$ (Interface B):} Participants refined outputs using \sysname{}'s auto-discovered sliders. Each task exposed between one and three discovered dimensions, and participants could also add a custom dimension. Although \sysname{} supports prompt-based editing, participants in this condition were instructed to rely on the discovered controls during the study tasks so that we could isolate the effect of slider-based steering.

\end{itemize}

Condition order was counterbalanced: 8 participants completed $C_{PROMPT}$ first and 8 completed $C_{STEER}$ first.
We chose a prompt-only baseline because it reflects the dominant way writers currently control general-purpose LLM writing tools: by issuing follow-up natural-language instructions. This main comparison allows us to test whether adding prompt-local direct-manipulation controls changes the authoring experience relative to current chat-based practice. It does not by itself isolate automatic discovery from other possible sources of slider controls, such as designer-authored, user-specified, or LLM-suggested dimensions; we return to this limitation in Section~\ref{sec:limitations}. 
In each condition, participants completed five revision tasks starting from provided drafts: four shared prompts and one participant-specific open-ended prompt. The four shared prompts covered two descriptive writing tasks and two workplace communication tasks. The participant-specific prompt allowed each participant to test the interfaces on a writing goal closer to their own needs, while holding that goal constant across both interfaces. For each prompt, participants revised the draft until they were satisfied with the result. Participants were asked to think aloud during the open-ended task.   

\begin{table}[t]
   \centering
   \begin{tabular}{p{0.1\linewidth} p{0.4\linewidth} p{0.3\linewidth}}           
   \toprule
   Prompt & Task prompt & Surfaced dimensions \\
   \midrule
   P1 & Describe the chaotic energy of a packed night market in a bustling city. & Vivid, Sensational, Atmosphere \\ 
   P2 & Describe a sudden thunderstorm hitting a quiet, open field during a hot afternoon. & Vividness, Sensory, Dissonance \\ 
   P3 & Write a brief, encouraging announcement message for a team channel celebrating a successful project launch. & Formality, Enthusiasm, Tone \\ 
   P4 & Write a brief, urgent Slack message asking teammates to hop onto a bridge call to fix a sudden production crash. & Urgency, Formality, Directness \\
  \bottomrule
  \end{tabular}
  \caption{Shared study prompts and surfaced dimensions.}
  \label{tab:study-prompts}
\end{table}
For the four shared prompts, all participants saw the same backend-discovered axes listed in Table~\ref{tab:study-prompts}, which held the discovered control set constant across participants. For participant-specific open-ended prompts, axes were generated for that participant's prompt.

Before interacting with either interface for a task, participants completed a pre-articulation step in which they stated in one or two sentences what they intended to change about the draft. This provided a baseline for interpreting whether \sysname{}'s discovered controls reflected changes participants had already articulated or instead helped them recognize dimensions they had not initially named. In $C_{STEER}$, participants completed this pre-articulation step before viewing the discovered slider labels. After the sliders were revealed, they were asked to comment on whether the surfaced dimensions matched, refined, or differed from the changes they had intended. 

\subsection{Measures and Analysis}

After each interface block, participants completed a questionnaire using 7-point Likert scales. Workload items asked about mental demand, physical demand, time pressure, perceived success, effort, and frustration. Additional interface items asked whether the interface made it easy to explore options, supported creativity, provided control over style, produced expected changes when adjusted, and helped participants match the text to what they had in mind. For \sysname{}, participants also rated whether slider labels accurately described the changes they made, whether sliders changed only what their labels described, whether the surfaced dimensions were things they wanted to adjust, and whether they would have thought to ask for the same style changes in a written prompt.

To complement subjective workload ratings, we summarized interaction logs from each task. Logs captured task duration, revision actions, diff-view use, and custom-dimension use. In $C_{PROMPT}$, revision actions corresponded to submitted natural-language reprompts; in $C_{STEER}$, they corresponded to committed slider adjustments. We use these process measures descriptively to contextualize perceived effort and to examine whether lower workload ratings were accompanied by differences in revision behavior.
 
After completing both conditions, participants answered forced-choice comparison questions and open-ended questions about what they liked, what felt limiting, and what features were missing. Because responses were collected on ordinal scales in a within-subjects design, we analyzed paired condition differences using exact Wilcoxon signed-rank tests and report medians, ranges, and exact $p$-values. For \sysname{}-specific items, we used exact one-sample Wilcoxon signed-rank tests against the neutral midpoint of 4. For forced-choice comparisons, we used exact binomial tests against chance. We treat the interaction logs as descriptive behavioral evidence rather than inferential outcomes.
 
\subsection{Results}

\subsubsection{RQ1: Effort Decreased Substantially While Perceived Success Remained Comparable}   

Participants reported substantially lower workload with \sysname{} than with the prompt-only baseline. Because responses were collected on 7-point ordinal scales in a within-subjects design, we analyzed paired condition differences using exact Wilcoxon signed-rank tests. Mental demand ratings in $C_{PROMPT}$ ranged from 4 to 6, with a median of 5, whereas ratings in $C_{STEER}$ ranged from 1 to 3, with a median of 2. All 16 participants rated mental demand lower in $C_{STEER}$ than in $C_{PROMPT}$ ($p < .001$).

Effort and frustration showed the same pattern. In $C_{PROMPT}$, 15 of 16 participants rated the effort required as 5--7, with a median of 6; in $C_{STEER}$, all 16 rated effort as 1--4, with a median of 2. Fifteen participants rated effort lower in $C_{STEER}$, with one tie ($p < .001$). Frustration ratings decreased from a median of 5 in $C_{PROMPT}$, with a range of 3--7, to a median of 2 in $C_{STEER}$, with a range of 1--3. All 16 participants rated frustration lower in $C_{STEER}$ ($p < .001$).

We also computed a workload composite by averaging mental demand, physical demand, time pressure, effort, and frustration. Internal consistency was acceptable in both conditions ($\alpha=.80$ for $C_{PROMPT}$ and $\alpha=.70$ for $C_{STEER}$). The composite showed the same pattern as the individual workload items: all 16 participants reported lower workload in $C_{STEER}$ than in $C_{PROMPT}$, with the median composite score decreasing from 4.8 to 1.9 ($p < .001$).

Interaction logs were consistent with these self-report results as shown in Table~\ref{tab:interaction-logs}. Participants spent less time per task in $C_{STEER}$ than in $C_{PROMPT}$: prompt-based tasks averaged 5.72 minutes per task (median = 5.70, range = 5.3--6.2), whereas slider-based tasks averaged 3.22 minutes per task (median = 3.20, range = 3.0--3.5). Averaging within participant, all 16 participants completed their $C_{STEER}$ tasks faster than their $C_{PROMPT}$ tasks, with a median reduction of 2.55 minutes per task. This reduction did not come from fewer revision actions. Participants made a number of iterations in both conditions: 3--4 reprompts per task in $C_{PROMPT}$ and 2--7 slider changes per task in $C_{STEER}$, with a median of 3.5 and 4.5 actions in the conditions, respectively. Thus, the behavioral logs suggest that \sysname{} reduced the cost of each revision step rather than simply reducing the number of attempts.

Perceived success remained high in both conditions. $C_{PROMPT}$ and $C_{STEER}$ both had a median success rating of 6, with no reliable paired difference ($p=.513$). Similarly, ratings for the item ``I could get the text to match what I had in mind'' were high in both conditions, with a median of 5 for $C_{PROMPT}$ and 6 for $C_{STEER}$; this difference was not reliable ($p=.480$). These results suggest that participants experienced \sysname{} as substantially less effortful without an observed decrease in perceived task success.

The post-study comparison questions echoed this effort reduction. When asked which interface required less effort, 14 of 16 participants selected Interface B, while 2 selected Interface A (exact binomial test against chance, $p=.004$). Overall preference favored Interface B descriptively, with 11 of 16 participants selecting Interface B and 5 selecting Interface A, although this forced-choice preference was not reliable under an exact binomial test ($p=.21$).     

\subsubsection{RQ2: Discovered Dimensions Supported Recognition Over Recall}

Participants strongly endorsed \sysname{}'s surfaced dimensions as relevant to their writing goals. On the item ``The style dimensions surfaced by the sliders were things I actually wanted to adjust in the writing tasks,'' all participants selected 5 or above on the 7-point agreement scale. The distribution was 2 ratings of 5, 6 ratings of 6, and 8 ratings of 7, yielding a median of 6.5. This was reliably above the neutral midpoint of 4 under an exact one-sample Wilcoxon signed-rank test ($p < .001$).       

At the same time, participants generally reported that they would not have thought to ask for those same changes in a written prompt. For the item ``I would have thought to ask for these same style changes in a written prompt,'' 15 of 16 participants selected 1, 2, or 3, with a median of 2. The distribution was 3 ratings of 1, 6 ratings of 2, 6 ratings of 3, and 1 rating of 4. No participant selected 5, 6, or 7. Ratings were significantly below the neutral midpoint ($p < .001$).

To directly capture recognition beyond recall, we computed a recognition-gap score for each participant by subtracting their rating for whether they would have asked for the same changes in a prompt from their rating for whether the surfaced dimensions were things they wanted to adjust. All 16 participants had positive recognition-gap scores, with a median gap of 4 points on the 7-point scale and a range of 1--6. This gap was reliably greater than zero under an exact one-sample Wilcoxon signed-rank test ($p < .001$).

The post-study forced-choice responses further support this recognition-over-recall interpretation. When asked which interface helped them notice style dimensions they had not thought of themselves, all 16 participants selected Interface B (exact binomial test against chance, $p < .001$). Participants' comments showed a spectrum of recognition beyond prior articulation. In some cases, \sysname{} gave a better label to an already-held intention: one participant who wanted a workplace message to have ``more seriousness'' recognized \emph{Formality} as ``what I meant,'' while another who wanted a message ``more straight to the point'' recognized \emph{Directness} as the desired control. In other cases, \sysname{} gave participants a vocabulary for intentions they could not yet name. One participant who initially said they were ``a little confused as to what to ask'' selected \emph{Vividness} and \emph{Sensory} as useful dimensions to try. Another, who wanted a storm description to feel more intense and uncomfortable, responded to the discovered \emph{Dissonance} control by saying, ``I couldn't think of how to describe it nor did I know to, but I like that slider.'' These cases suggest that discovered controls supported both relabeling of vague intentions and recognition of useful dimensions participants had not formulated as prompt instructions.

\subsubsection{RQ3: \sysname{} Slider Controls Were Useful, But Participants Wanted Hybrid Control}

Participants generally found \sysname{}'s controls interpretable and predictable, but their responses also revealed that slider steering is best understood as one part of a hybrid authoring workflow. In the study, $C_{STEER}$ constrained participants to focus on discovered sliders so that we could isolate the effect of slider-based steering from open-ended prompting. Thus, the tradeoffs reported here reflect the evaluated study condition, not a limitation of the full \sysname{} interface, which is designed to support both prompting and slider manipulation.

On the item ``The slider labels accurately described the changes the sliders made,'' all ratings were at or above the midpoint: 2 participants selected 4, 7 selected 5, 5 selected 6, and 2 selected 7, with a median of 5. Ratings were reliably above the neutral midpoint under an exact one-sample Wilcoxon signed-rank test ($p < .001$). Participants also found slider adjustments reasonably predictable. For ``When I made an adjustment, the text changed in the way I expected,'' all $C_{STEER}$ ratings were 4 or above, with a median of 5; this was also reliably above the midpoint ($p < .001$).

However, participants were more moderate about whether sliders changed only the labeled attribute. For ``The sliders changed only what their labels described,'' ratings ranged from 3 to 7, with a median of 4.5. This item was not reliably above the midpoint ($p=.074$), suggesting that participants perceived some entanglement or side effects in the discovered controls, even when they found the labels broadly useful.

Interaction logs suggest that participants used the comparison affordances to manage this uncertainty. Thirteen of 16 participants used the diff view during $C_{STEER}$ tasks, and 9 kept it open throughout their slider-based tasks. This suggests that participants did not treat discovered labels as self-explanatory; they actively inspected the textual consequences of steering. Five of 16 participants also used the custom-dimension feature during the study tasks, indicating that participants sometimes sought stylistic controls beyond the dimensions initially surfaced by the system.

The comparison between interfaces showed an important distinction between effort reduction and broader authoring control. As reported above, 14 of 16 participants selected Interface B as requiring less effort (exact binomial test against chance, $p=.004$). Overall preference favored Interface B descriptively, with 11 of 16 participants selecting Interface B and 5 selecting Interface A, but this forced-choice preference was not reliable under an exact binomial test ($p=.21$). Perceived control was divided: 9 of 16 participants selected Interface B as making them feel more in control of the text, while 7 selected Interface A ($p=.804$). Participants were also evenly split on which interface they would rather use for their own writing, with 8 selecting Interface A and 8 selecting Interface B ($p=1.00$). These results suggest that sliders reduced the effort of available adjustments, but participants still valued the expressive flexibility associated with natural-language prompting.

Open-ended responses help explain this tension. Participants who preferred Interface B emphasized reduced cognitive effort and the usefulness of automatically surfaced dimensions. One participant wrote that changes that would take ``multiple prompts on Interface A would take little effort to achieve on Interface B because of the sliders.'' Another wrote that they liked not needing ``to think about the dimensions to control'' because they could ``just move them through a slider.'' A third participant described \sysname{} as creating ``little to no mental strain'' because the relevant dimensions surfaced automatically.

Participants who preferred Interface A emphasized the expressive freedom of natural-language prompting. One participant wrote, ``I have the freedom to explain what exactly I want,'' while another wrote, ``It is up to me what I want to change.'' Several participants across both preference groups wanted to move more fluidly between slider steering and prompting during the study tasks. Because the full \sysname{} prototype supports prompt-based editing, but $C_{STEER}$ constrained participants to focus on sliders to isolate the effect of discovered controls, these comments should be interpreted as support for the intended hybrid workflow rather than as a limitation of the system architecture. For example, one participant wrote that Interface B would be better if they had ``the option to reprompt and use the sliders,'' and another wrote that they would like ``a mix of Interface A and B'' where they could also reprompt when needed. In this sense, participants' requests point toward the interaction model \sysname{} is designed to support: language for open-ended, structural, or factual changes, and discovered sliders for rapid recognition-based adjustment of stylistic dimensions.

\subsubsection{Supplementary Control-Source Evaluation}
To probe whether \sysname{}'s PCA-discovered controls differed from controls a general-purpose LLM could propose, we ran a small blinded label-only comparison with eight additional participants. For each of the four shared prompts, participants saw two unlabeled three-control sets: the PCA-discovered controls used in \sysname{} and a GPT5.6-luna-generated set of stylistic, tonal, or rhetorical controls for the same prompt \cite{openai_2026_luna}. GPT5.6-luna~\cite{openai_2026_luna} is a publicly available commercial chat model that was accessible on a free tier at the time of the study. Descriptively, across 32 set-level judgments, PCA-discovered sets were selected 21 times (65.6\%) and GPT5.6-luna-generated sets 11 times (34.4\%). Individual dimensions from both sources were often judged useful (PCA: 78/96, 81.3\%; GPT5.6-luna: 81/96, 84.4\%), but PCA dimensions were more often useful-but-not-anticipated (28/96, 29.2\%) than GPT5.6-luna dimensions (16/96, 16.7\%). Excluding labels shared by both sources, source-unique PCA dimensions were useful-but-not-anticipated in 28/64 judgments (43.8\%), compared with 16/64 (25.0\%) for source-unique GPT5.6-luna dimensions. Because this was a label-only comparison rather than an interactive slider baseline, we treat it as exploratory.

\begin{table}[t]
\centering
\begin{tabular}{lcc}
\toprule
Measure & $C_{PROMPT}$ & $C_{STEER}$ \\
\midrule
Mean task duration & 5.72 min & 3.22 min \\
Median task duration & 5.70 min & 3.20 min \\
Task duration range & 5.3--6.2 min & 3.0--3.5 min \\
Median revision actions & 3.5 reprompts & 4.5 slider changes \\
Revision action range & 3--4 & 2--7 \\
Diff view used & 11/16 & 13/16 participants \\
Diff view kept open & 6/16 & 9/16 participants \\
Custom dimensions used & -- & 5/16 participants \\
\bottomrule
\end{tabular}
\caption{Descriptive interaction-log summaries across participants and tasks.}
\label{tab:interaction-logs}
\end{table}

\section{Discussion}

Our study suggests that prompt-local steering controls are most valuable not because they replace language, but because they change what users must do cognitively during revision. In the prompt-only baseline, participants had to inspect a draft, decide what was wrong, translate that judgment into a prompt, and then evaluate whether the resulting revision matched their intent. In \sysname{}, the system instead externalized candidate dimensions of variation as manipulable controls. This shift reduced perceived mental demand, effort, frustration, and task time while preserving comparable perceived success. At the same time, participants' divided responses on control and own-writing preference complicate a simple ``sliders are better than prompting'' interpretation. Below, we discuss four implications for LLM authoring interfaces.

\subsection{From Prompt Engineering to Recognition-Based Control}

Our results reinforce prior work showing that effective LLM use depends not only on model capability, but also on whether users can express what they want~\cite{zamfirescu2023johnny,subramonyam2024envisioning}. In writing, that expression problem is often stylistic: users may sense that a draft feels wrong without knowing how to name the desired change. \sysname{} addresses this problem by surfacing prompt-local dimensions of generation variation as candidate controls.

Participants' responses support this recognition-over-recall framing. All 16 participants rated the surfaced dimensions as things they wanted to adjust, while 15 of 16 disagreed that they would have thought to ask for the same changes in a written prompt. This suggests that the relevant preference was often recognized after being made visible rather than recalled in advance. The result aligns with work on constructed preferences and interactive elicitation, which argues that preferences are often assembled through interaction with available options~\cite{slovic1995construction,li2023eliciting}.

This finding also clarifies the value of prompt-locality. Unlike fixed tone controls or widgets derived from already-stated preferences, \sysname{} surfaces dimensions from the model's own variation for the current prompt. A workplace announcement may expose \emph{Formality} or \emph{Enthusiasm}, while a descriptive task may expose \emph{Vividness}, \emph{Sensory} detail, or \emph{Atmosphere}. This makes the controls adaptive to the current writing situation. However, prompt-locality is also a constraint: a useful dimension for one prompt may not appear for another, and the same label may behave differently across tasks. Interfaces should therefore present prompt-local controls as contextual hypotheses rather than universal model attributes.

\subsection{Effort Reduction and Perceived Control Are Different Outcomes}

The study separates two outcomes that are often conflated: reducing effort and increasing perceived control. \sysname{} clearly reduced effort. Participants reported lower mental demand, effort, and frustration; interaction logs showed shorter task durations; and 14 of 16 participants selected Interface B as requiring less effort. Qualitative comments echoed this result: participants described the sliders as reducing the need to think through and phrase each desired change.

However, this effort reduction did not translate into uniform preference for \sysname{} as the more controllable system. Perceived control was divided, with 9 of 16 participants selecting Interface B and 7 selecting Interface A. Participants were also evenly split on which interface they would rather use for their own writing. These results suggest that participants distinguished between low-effort adjustment and open-ended expressive freedom.

This distinction is important for LLM interface design. Sliders make available adjustments fast, visible, and reversible, consistent with classic direct-manipulation principles~\cite{shneiderman1983direct,hutchins1985direct}. But language remains better suited for structural, factual, idiosyncratic, or open-ended changes. Several participants explicitly asked for a mix of both interfaces. Because the full \sysname{} prototype supports prompting, but the study constrained $C_{STEER}$ to focus on sliders, these comments should be interpreted as support for the intended hybrid workflow rather than as evidence that prompting should be removed. Future authoring systems should let users reprompt while preserving useful slider states, compose prompt edits with active steering settings, and compare versions across both modes.

\subsection{Discovered Controls Must Be Inspectable and Repairable}

A central design choice in \sysname{} is to treat discovered axes as interface hypotheses rather than precise semantic knobs. Participants generally found slider labels descriptive and adjustments predictable, but they were more moderate about whether sliders changed only the labeled attribute. This reflects a known limitation of activation-space steering: useful linear directions can still be entangled with length, specificity, emotional tone, formatting, or other properties~\cite{park2024linear,engels2025linear}.

Our findings show how this technical limitation appears at the interface level. Users can often make sense of a discovered control, but they still need ways to inspect what it actually changed. Thirteen of 16 participants used the diff view during $C_{STEER}$ tasks, and 9 kept it open throughout their slider-based tasks. This suggests that participants did not treat labels as self-explanatory; they actively checked the textual consequences of steering. Because \sysname{} performs draft-level rather than span-level steering, this comparison support is especially important: a single slider movement can subtly alter multiple parts of the output.
 
Repair mechanisms were also important. Five participants used the custom-dimension feature, indicating that discovered controls sometimes failed to cover the dimension a participant wanted. Pole previews, variance tags, diff views, reset controls, and custom CAA-based dimensions are therefore not secondary interface features. They are what make imperfect machine-discovered controls usable. More broadly, controllable generation systems should combine steering mechanisms with accountability mechanisms that let users see, test, reverse, and repair the effects of control.

\subsection{Implications for Design and Evaluation}

Our findings suggest several implications for controllable LLM authoring systems. First, interfaces should support recognition before articulation. Users may recognize useful stylistic directions only after they are surfaced, so the prompt should not be treated as the only site where intent can be expressed. Second, discovered controls should be presented as provisional and inspectable. Labels approximate patterns of model variation; they should be paired with examples, diffs, reset controls, and other feedback that lets users judge whether the control is useful for the current task. Third, prompting should remain available as an expressive escape hatch. Sliders reduce effort for available dimensions, but users still need language for precise, structural, factual, or idiosyncratic revisions. Fourth, interfaces should communicate prompt-locality: a control discovered for one prompt should not be implied to be a stable global attribute of the model.

These findings also have implications for evaluation. A single preference or output-quality measure would miss the tradeoff we observed. \sysname{} reduced effort and supported recognition, yet did not dominate prompt-only interaction on perceived control or own-writing preference. Future evaluations of controllable LLM interfaces should therefore measure effort, recognition, interpretability, perceived control, recovery from mismatch, and final output quality separately. Our supplementary label-only comparison with GPT5.6-luna-generated controls provides an initial check against one publicly available LLM-suggested control source, suggesting that PCA-discovered dimensions may contribute more useful-but-not-anticipated controls. However, future evaluations should still include full interactive component baselines when possible, such as fixed designer-authored sliders, user-specified sliders, LLM-suggested sliders, or Malleable Prompting-style widgets. Thus, the supplementary comparison narrows but does not eliminate the need for component-level interactive evaluations. 
 
Overall, \sysname{} illustrates how prompt-local steering controls can reduce the burden of articulating every stylistic revision in advance. The goal is not to replace prompting, but to combine language with direct manipulation so that writers can discover, inspect, and steer useful dimensions of variation as they work. 

\section{Limitations and Future Work}
\label{sec:limitations}

This study has several limitations. First, it involved 16 participants and short-form writing tasks, so the findings should be interpreted as evidence about interaction patterns rather than broad claims about all LLM authoring contexts. Future work should evaluate longer documents, professional writing settings, collaborative writing, and a wider range of genres.

Second, the $C_{STEER}$ condition constrained participants to focus on sliders even though the full \sysname{} prototype supports prompt-based editing. This design isolated slider-based steering for the study, but it made the condition less naturalistic and likely reduced perceived open-ended control relative to the intended hybrid prompt-and-slider workflow.   

Third, the prototype relied on a live Colab-hosted backend exposed through LocalTunnel. This made the study more representative of actual end-to-end interaction than a precomputed demo, but it also introduced deployment constraints: GPU availability, Colab session stability, tunnel reliability, and backend queueing could affect latency and responsiveness. Future deployments should use more stable serving infrastructure and report end-to-end latency for generation, discovery, labeling, and steering separately.

Fourth, the main study compared \sysname{}'s discovered sliders against a matched prompt-only baseline. We chose this baseline because prompt-only chat remains the dominant interaction pattern for LLM writing, and our goal was to test whether adding prompt-local direct-manipulation controls changes that experience. Our supplementary control-source evaluation provides an initial blinded, label-only comparison with GPT5.6-luna-generated controls, but it did not test a full interactive interface with the same slider UI and different control sources. A stronger component-level evaluation would compare \sysname{} against interactive baselines using fixed designer-authored dimensions, user-specified dimensions, LLM-suggested dimensions, or Malleable Prompting-style widgets. Our results therefore support claims about \sysname{} as an integrated discovered-control authoring interface relative to current prompt-only practice, while leaving open how much of the benefit comes specifically from the discovery mechanism versus slider-based direct manipulation more broadly.

Finally, the current prototype uses PCA-discovered directions from one target model and applies draft-level steering at a fixed layer. PCA identifies prominent directions of activation variance, not guaranteed disentangled semantic factors, and participants' ratings suggest that sliders sometimes affected properties beyond their labels. Future work should evaluate additional models, compare PCA with other steering-direction discovery methods, audit label and polarity accuracy with independent raters, and assess output quality, content preservation, and long-range coherence with blinded evaluations.

\section{Conclusion}
We presented \sysname{}, an LLM authoring interface that turns prompt-local variation in model generations into interactive steering controls. Rather than requiring writers to fully specify stylistic preferences through repeated prompting, \sysname{} surfaces candidate dimensions that users can inspect, manipulate, reset, and supplement.
 
In a within-subjects study with 16 participants, \sysname{} reduced perceived mental demand, effort, and frustration relative to a matched prompt-only baseline while preserving comparable perceived success. Participants also identified the surfaced dimensions as changes they wanted, even though most reported they would not have thought to ask for those changes in a prompt. Interaction logs further suggested that slider-based revision reduced the time cost of each iteration.

At the same time, participants continued to value natural-language prompting for open-ended control. The broader design implication is therefore not that sliders should replace prompting, but that LLM authoring interfaces should combine the flexibility of language with the immediacy of direct manipulation. \sysname{} illustrates one path toward hybrid systems that help writers discover, inspect, and steer useful dimensions of variation that might otherwise remain unarticulated.

%%
%% The acknowledgments section is defined using the "acks" environment
%% (and NOT an unnumbered section). This ensures the proper
%% identification of the section in the article metadata, and the
%% consistent spelling of the heading.
\begin{acks}
Acknowledgments are omitted for anonymous review.
\end{acks}

\section*{Ethics and Privacy Statement}

This work studies interfaces for controllable LLM-based writing. Participants completed writing, revision, and control-preference tasks and provided survey responses and open-ended feedback about their experience. We report results in aggregate and use anonymized quotes; no participant names or identifying information are included in the paper.

The system is intended to increase user agency by making stylistic variation inspectable and adjustable. However, more controllable style steering could also be misused to produce manipulative, deceptive, or impersonated text. We therefore frame \sysname{} as an authoring and revision aid, emphasize transparency through labels, pole previews, diff views, and reset controls, and discuss the need for future deployments to include safeguards appropriate to their writing context.     

The study evaluated a bounded research prototype with fixed study tasks and constrained steering coefficients. Our findings should therefore be interpreted as evidence about interaction design and user experience, not as an endorsement of unrestricted style manipulation in high-stakes or deceptive communication settings.

%%
%% The next two lines define the bibliography style to be used, and
%% the bibliography file.
\bibliographystyle{ACM-Reference-Format}
\bibliography{references}

%%
%% If your work has an appendix, this is the place to put it.

\appendix

\section{Technical Details of Dimension Discovery and Steering}
\label{app:technical}

This appendix provides additional implementation details for the prompt-local dimension discovery, labeling, and steering pipeline described in Section~\ref{sec:Implementation}.

\subsection{Sampling Parameters}
\label{app:sampling-layer}

For reproducibility, the discovery pipeline sampled $N=30$ completions per prompt using nucleus sampling with temperature $T=0.9$, top-$p=0.95$, and a maximum length of 500 new tokens. 

\subsection{Prior validation of axis discovery.}
Prior work evaluated the same sampling-based PCA discovery procedure on Qwen3.5-4B over 20 prompts with eight annotators, who produced 245 free-text ``magic slider'' requests~\cite{mallavarapu2026sampling}. Top-two axes achieved 72.8\% precision and 43.6\% macro-recall at $\tau=0.65$; because only two axes were evaluated, the estimated recall ceiling was 51.7\%. Pole generations had median validity 4/5, 75.6\% top-two-box accuracy, 57.5\% exact agreement, and 90.9\% adjacent agreement. Across nine seeds, axes had 82.5\% top-two rediscovery; a text-only PCA baseline reached 38.4\% precision and 21.2\% recall.

\subsection{Layer Sweep Calibration}                  
\label{app:layer-sweep}

To choose a fixed intervention layer for the prototype, we ran a pilot sweep over candidate transformer layers $\{12,16,18,21,24,25\}$. The sweep used four calibration prompts spanning descriptive writing and workplace communication tasks. For each layer and prompt, we ran the discovery pipeline, retained the top three PCA axes, and generated paired steered outputs at $\alpha=-1.5$ and $\alpha=+1.5$, yielding 72 layer--prompt--axis configurations.

We evaluated these configurations with \texttt{Meta-Llama-3-8B-Instruct} as an automated judge. For each configuration, the judge was shown the original prompt, the discovered axis label, and the two steered outputs. The judge assessed whether the two outputs formed a coherent contrast along the labeled dimension, whether each output preserved the original task, whether the outputs remained fluent, and whether they contained major artifacts such as reasoning leakage, prompt continuation, unnecessary constraint insertion, multiple-option answers when not requested, or task failure.            

The sweep did not identify a single layer that was uniformly optimal across all prompts and axes. Instead, usable steering was concentrated in a mid-to-late band: layers 18--24 produced the most consistent trade-off between visible steering movement and preservation of fluent, task-adherent outputs. Earlier and later candidates were more prone to either weaker steering or output instability under the same coefficient range. We therefore selected layer 21 as a representative fixed layer within this high-performing band, rather than optimizing the layer separately for each prompt or dimension. We treat this as an implementation calibration for Qwen3.5-4B, not as evidence that layer 21 is universally optimal across models, prompts, or steering methods.

\subsection{Dimension Labeling and Polarity Alignment}
\label{app:labeling}

Because the sign of a PCA direction is arbitrary, each discovered direction must be grounded in text examples before it can be presented as an interpretable slider. For each discovered principal component $\mathbf{v}_k$, we projected the $N$ sampled generations onto the axis and selected representative examples from the negative and positive ends of the projection distribution. These examples were passed to \texttt{Meta-Llama-3-8B-Instruct}~\cite{llama3herdofmodels} as pole examples for labeling and polarity alignment.

\begin{quote}
\small
\textbf{Labeling prompt template:}

You are an expert linguistic annotator evaluating text outputs from a language model. The following two sets of completions were generated from the same initial prompt and come from opposite ends of one discovered variation axis.

   \textbf{Negative pole examples:}
   1. [negative pole sample 1]                           
   2. [negative pole sample 2]
   3. [negative pole sample 3]

   \textbf{Positive pole examples:}
   1. [positive pole sample 1]
   2. [positive pole sample 2]
   3. [positive pole sample 3]

   Compare the two sets. Identify the most salient stylistic, tonal, or rhetorical property that differs between the negative pole and the positive pole.

   Return JSON only in the following format:

   \begin{verbatim}
   {
     "label": "<one- or two-word label>",
     "positive_pole_has_more_label": <true or false>,
     "coherence": <numeric score>,
     "rationale": "<one short sentence>"
   }
   \end{verbatim}

   The \texttt{label} should be a concise one- or two-word name for the property. Set \texttt{positive\_pole\_has\_more\_label} to \texttt{true} if the positive pole examples exhibit more of the labeled property than the negative pole examples, and \texttt{false} otherwise. The \texttt{coherence} field should indicate how coherent or interpretable the contrast is. The \texttt{rationale} should be one short sentence.
\end{quote}

The \texttt{positive\_pole\_has\_more\_label} field was used for polarity alignment. If the judge returned \texttt{false}, we multiplied the PCA direction by $-1$ and swapped the displayed pole examples. After this alignment step, rightward slider movement consistently corresponded to increasing the labeled property.

\subsection{Interactive Steering Mechanism Calibration}
\label{app:steering-calibration}
To determine the usable parameter ranges for activation-space interventions, we calibrated the steering coefficient through pilot sweeps over $\alpha \in \{-3, -2.5, -2, -1.5, -1, -0.75, -0.5, 0, 0.5, 0.75, 1, 1.5, 2, 2.5, 3\}$. Based on qualitative inspection of output fluency and task adherence, we established $\alpha \in [-1.5, 1.5]$ as the conservative slider boundary used in the live prototype.

\subsection{Supplementary Control-Source Evaluation}
\label{app:control-source}

We conducted a small blinded supplementary evaluation comparing PCA-discovered controls with GPT5.6-luna-generated controls for the four shared study prompts. GPT5.6-luna is a publicly available commercial chat model accessible on a free tier at the time of the study. It was used only to generate the supplementary LLM-suggested control baseline and was not part of the \sysname{} discovery, labeling, or steering pipeline. Eight additional participants completed all four prompt evaluations. We did not collect demographic information for this supplementary evaluation.

For each prompt, participants saw two unlabeled three-control sets, presented as Set A and Set B. One set contained the PCA-discovered controls used in \sysname{}; the other contained three stylistic, tonal, or rhetorical controls generated by GPT5.6-luna for the same prompt. Source position varied across tasks: for Tasks 1 and 3, Set A was PCA and Set B was GPT5.6-luna; for Tasks 2 and 4, Set A was GPT5.6-luna and Set B was PCA.

The GPT5.6-luna controls were generated with a fixed prompt asking the model to propose exactly three useful controls for revising the writing task:

\begin{quote}
\small
Given the following writing task, propose exactly three useful controls that a writer might want while revising the model's output. Each control should be a one- or two-word stylistic, tonal, or rhetorical dimension. Avoid content-specific nouns. Avoid redundant controls. Return only the three labels as a comma-separated list.

Writing task: [task]
\end{quote}

Participants first selected which complete set of controls they would rather have while revising the text. They then selected which individual dimensions they considered useful, and which dimensions they believed they would have independently thought to ask an AI writing assistant to adjust before seeing the list. For prompts where both sources produced the same label, the shared dimension was displayed once in the individual-dimension questions but counted once for each source in source-attributed summaries.

Across 32 set-level judgments, PCA-discovered sets were selected 21 times (65.6\%) and GPT5.6-luna-generated sets 11 times (34.4\%); no participant selected ``no preference.'' Five of eight participants selected the PCA set more often, one selected the GPT5.6-luna set more often, and two selected the two sources equally often.

At the individual-dimension level, PCA-discovered dimensions were selected as useful in 78 of 96 source-attributed judgments (81.3\%), while GPT5.6-luna-generated dimensions were selected in 81 of 96 judgments (84.4\%). PCA-discovered dimensions were independently anticipated in 50 of 96 judgments (52.1\%), compared with 65 of 96 judgments (67.7\%) for GPT5.6-luna-generated dimensions. Consequently, PCA-discovered dimensions were more often useful-but-not-anticipated: 28 of 96 judgments (29.2\%) versus 16 of 96 judgments (16.7\%) for GPT5.6-luna-generated dimensions.

Because Tasks 3 and 4 included controls shared by both methods, we also computed a source-unique summary excluding shared labels. Among 64 source-unique judgments per source, PCA-discovered dimensions were useful in 46 judgments (71.9\%), independently anticipated in 18 (28.1\%), and useful-but-not-anticipated in 28 (43.8\%). GPT5.6-luna-generated dimensions were useful in 49 judgments (76.6\%), independently anticipated in 33 (51.6\%), and useful-but-not-anticipated in 16 (25.0\%).

These results suggest a difference in the character of the two control sources. GPT5.6-luna-generated controls were slightly more often judged useful overall and more often independently anticipated, suggesting that they captured familiar revision dimensions. PCA-discovered controls were more often useful-but-not-anticipated, suggesting that prompt-local activation variation can surface less obvious but still valuable controls. Because this was a small label-only evaluation rather than an interactive slider condition, we treat it as exploratory.                                                                     

\begin{table}[h]
\centering
\small
\begin{tabular}{p{0.12\linewidth} p{0.36\linewidth} p{0.36\linewidth}}
\toprule
Prompt & PCA-discovered controls & GPT5.6-luna-generated controls \\
\midrule
P1 & Vivid; Sensational; Atmosphere & Immersiveness; Sensory Intensity; Pacing \\
P2 & Vividness; Sensory; Dissonance & Dramatic Contrast; Sensory Detail; Atmospheric Tension \\
P3 & Formality; Enthusiasm; Tone & Warmth; Enthusiasm; Formality \\
P4 & Urgency; Formality; Directness & Directness; Reassurance; Urgency \\
\bottomrule
\end{tabular}
\caption{Control sets shown in the supplementary control-source evaluation.}
\label{tab:control-source-sets}
\end{table}

\end{document}